\documentclass[12pt,a4paper]{article}
\usepackage{amsmath,amssymb}
\usepackage{graphicx}
\begin{document}
\newcounter{Series}
\setcounter{Series}{2}
\title{
Revival of Single-Particle Transport Theory 
for the Normal State of High-$T_c$ Superconductors: \\
\Roman{Series}. 
Vertex Correction }

\author{
O. Narikiyo
\footnote{
Department of Physics, 
Kyushu University, 
Fukuoka 812-8581, 
Japan}
}

\date{
(Jan. 25, 2013)
}

\maketitle
\begin{abstract}
The vertex correction for the electric current 
is discussed on the basis of the Fermi-liquid theory. 
It does not alter the qualitative description 
of the electric transport 
by the relaxation-time approximation 
in the case of the normal state of high-$T_c$ superconductors. 
The failure of the transport theory 
employing the fluctuation-exchange (FLEX) approximation 
is pointed out. 
It becomes manifest 
by considering the Ward identity for the fluctuation mode. 
\end{abstract}
\vskip 30pt

In the previous Short Note, arXiv:1204.5300v3, 
the DC Hall conductivity is discussed 
on the basis of the relaxation-time approximation. 
Such an approximation is enough 
to explain the qualitative features of the measured DC Hall conductivity 
in the normal state of high-$T_c$ superconductors. 
The vertex correction is irrelevant. 
The reason is discussed in this Short Note. 
Especially the failure\footnote{
I have enforced the critiques 
reported in cond-mat/0006028v1, 0012505v1, 0103436v1. } 
of the transport theory 
based on the fluctuation-exchange (FLEX) approximation is pointed out. 
In the next Short Note I shall discuss the AC Hall conductivity. 

The discussion on the vertex correction 
for the electric current based on the Fermi-liquid theory 
is reviewed in Chap.\,VIII of \cite{AGD}. 
More explicit diagrammatic representation is given in \cite{YY}. 
The following discussion is based on these~\cite{AGD,YY}. 

We start from the self-energy for electrons shown in Fig\,1 
where the electrons with ${\bf k}$ and ${\bf k}\!-\!{\bf q}$ 
carry spin $\sigma$ and 
the others with ${\bf k}'$ and ${\bf k}'\!+\!{\bf q}$ 
carry spin $\sigma'$. 
The diagrammatic representation for the electric current vertex 
$\Lambda_\mu({\bf k})$, where $\mu=x,y,z$, is also given in Fig.\,1. 
The vertex correction consistent with the Ward identity 
is obtained as Fig.\,2 
by introducing the coupling to external field 
to one of the internal lines of the self-energy. 
Thus the integral equation to determine $\Lambda_\mu({\bf k})$ is 
\begin{equation}
\Lambda_\mu({\bf k}) = J_\mu({\bf k}) 
+ \Lambda_\mu^{(a)}({\bf k}) 
+ \Lambda_\mu^{(b)}({\bf k}) 
+ \Lambda_\mu^{(c)}({\bf k}), 
\label{int} 
\end{equation}
where $J_\mu({\bf k})$ is the bare vertex, 
$\Lambda_\mu^{(a)}({\bf k})$ is the integral 
depicted in Fig.\,2-(a) and so on. 
This equation is rewritten as\footnote{
See Eqs.\,(6$\cdot$13) and (6$\cdot$17) in \cite{YY}. } 
\begin{align}
0 = 
J_\mu({\bf k}) + \sum_{{\bf k}'} \sum_{{\bf q}} & 
\Delta_0({\bf k},{\bf k}';{\bf k}'\!+\!{\bf q},{\bf k}\!-\!{\bf q}) 
\nonumber \\ 
& \times 
\Big[ \Phi_\mu({\bf k}\!-\!{\bf q}) + \Phi_\mu({\bf k}'\!+\!{\bf q}) 
- \Phi_\mu({\bf k}') - \Phi_\mu({\bf k}) \Big], 
\label{coll} 
\end{align}
at low temperatures 
where the Fermi degeneracy becomes a strong constraint. 
This equation is consistent with the collision term\footnote{
See Eqs.\,(40.2) and (40.6) in \cite{AGD}. } 
in the Boltzmann equation. 
Such a form is the consequence 
of the Fermi statistics (the Pauli principle). 

On the other hand, 
in the FLEX approximation~\cite{Kon} 
the Aslamazov-Larkin processes\footnote{
See Fig.\,13 in \cite{Kon}.} 
corresponding to 
$\Lambda_\mu^{(b)}({\bf k})$ and $\Lambda_\mu^{(c)}({\bf k})$ vanish 
and only the Maki-Thompson process 
corresponding to $\Lambda_\mu^{(a)}({\bf k})$ 
contributes to the integral equation. 
Thus this approximation does not lead to the vertex obeying Eq.\,(2) 
and is not applicable to the system with the Fermi degeneracy. 

The reason why the Aslamazov-Larkin processes vanish 
has nothing to do with the Umklapp scattering\footnote{
The role of the Umklapp scattering is stressed in \cite{Kon}. }. 
They vanish for any charge-neutral fluctuation mode. 
It is rigorously shown by the Ward identity\footnote{
See arXiv:1212.6484v1 and references therein. }. 

A correct transport theory for electrons at low temperatures\footnote{
At low temperatures 
the transport theory should based on quasi-particles 
obeying the Pauli principle. 
On the other hand, 
the superconducting fluctuation transport theory 
is formulated in high-temperature limit. } 
should obey the Pauli principle. 
However, the replacement of the renormalized interaction 
(depicted by the square) 
by the fluctuation mode in the FLEX approximation 
breaks the Pauli principle~\cite{BW,DHS,VT}. 
Only internal consistency is guaranteed by the FLEX approximation 
but the difference between the exact result and the approximation 
cannot be estimated in its framework. 

As discussed in \S 39 of \cite{AGD} 
the vertex correction for elastic scatterings 
removes the contribution of the forward scattering\footnote{ 
It is represented by the factor $(1-\cos\theta)$ 
in Eq.\,(39.17) of \cite{AGD}. }. 
This effect is irrelevant\footnote{
See, for example, B. P. Stojkvi\'c and D. Pines: 
Phys. Rev. B {\bf 55}, 8576 (1997). } to the qualitative discussion 
in the case of high-$T_c$ superconductors 
where the forward scattering is negligible. 

Consequently 
the relaxation-time approximation is enough 
to describe the qualitative features of the transport 
in the normal state of high-$T_c$ superconductors. 

In the previous Short Note, arXiv:1204.5300v3, 
it is clarified that the qualitatively correct self-energy is necessary 
to obtain the transport coefficients consistent with experiments. 
However, such a self-energy\footnote{
The vertex correction 
for the interaction between the electron and the fluctuation mode 
is neglected in the self-energy of the FLEX approximation. 
While in the case of electron-phonon interaction 
Migdal's theorem guarantees that the vertex correction is negligible, 
in the case of electron-fluctuation interaction 
such a theorem is absent 
so that the vertex correction is not negligible. } 
cannot be obtained~\cite{VT} by the FLEX approximation. 

\vskip 30pt



\newpage 

\begin{figure}[htbp]
\begin{center}
\includegraphics[width=9.4cm]{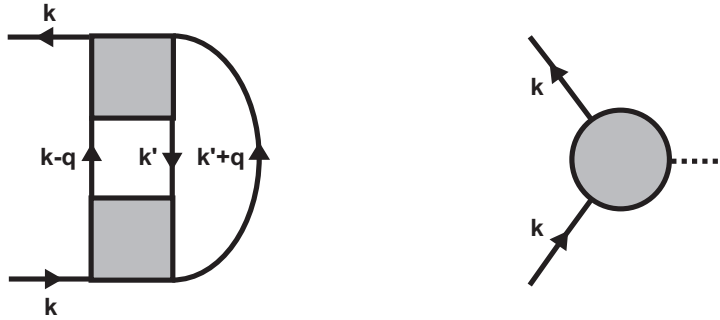}
\vskip 2mm
\caption{(Left) Self-energy. (Right) Current vertex.}
\label{fig:self-energy}
\end{center}
\end{figure}
\vskip 15mm
\begin{figure}[htbp]
\begin{center}
\includegraphics[width=16.0cm]{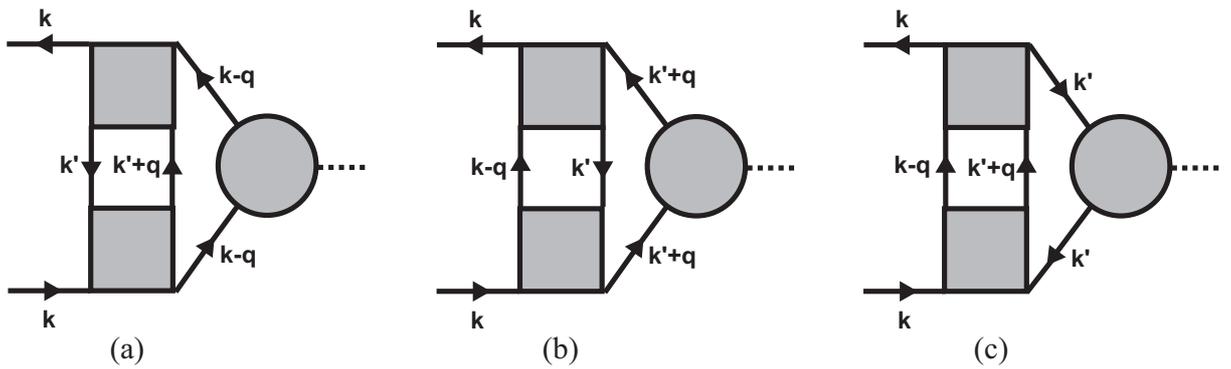}
\vskip 2mm
\caption{Vertex corrections.}
\label{fig:vertex}
\end{center}
\end{figure}

\end{document}